# Persistence of temperature and precipitation: from local to global anomalies

Jouni J. Takalo

University of Oulu and University of Jyväskylä, Finland

## Abstract

Using detrended fluctuation analysis (DFA) we find that all continents are persistent in temperature. The scaling exponents of the southern hemisphere (SH) continents, i.e., South America (0.77) and Oceania (0.72) are somewhat higher than scaling exponents of Europe (0.70), Asia (0.69) and North America (0.64), but the scaling of Africa is by far the highest (0.86). The reason for this is the location of Africa near the equator.

The scaling exponents of the precipitation are much smaller, i.e. between 0.55 (Europe) and 0.68 (North America). The scaling exponent of Europe is near that of the random noise (0.5), while the other continents are slightly persistent in precipitation. We also show that the persistence disappears in all time series when shuffling the data randomly, showing that persistence is not an intrinsic property of the estimator.

We find that the monthly temperature is the more persistent the more wide area is analyzed. The persistence of precipitation is also increasing as a function of area, although not so clearly. Furthermore, our analysis shows that the persistence of the temperature decreases poleward in both hemispheres. The persistence of precipitation is also highest at equatorial zones, and decreases poleward, but is less dependent on the latitude than temperature. It seems that the persistence of the precipitation is more dependent on the corresponding climate type than the persistence of the temperature.

Keywords: Temperature; Precipitation; Persistence; Scaling; Detrended.

## 1. Introduction

Several studies have been published on the long-range temporal correlations of temperature, usually by studying the Hurst exponent (H) of the temperature anomalies. H is also called scaling exponent, $\alpha$, which we will use in this study. Scaling exponent is a measure of the rate of the statistical dependence of the points as a function of their distance in the time series. The value $\alpha = 0.5$ is the scaling exponent of random fractional Brownian noise. If $\alpha > 0.5$ for the time series, we say that it is persistent, and has longer-rate memory, while the value $\alpha < 0.5$ means anti-persistence in the behavior of the time series. The majority of the studies use local temperature records, and give exponents around 0.7 varying between 0.5-0.8 [Bodri, 1994; Pelletier and Turcotte, 1997; Koscielny-Bunde et al., 1998; Mann, Bradley and Hughes, 1998; Eichner et al., 2003, Fraedrich and Blender, 2003; Monetti, Havlin and Bunde, 2003; Bunde et al., 2004; Király, Bartos and Jánosi, 2006; Zhang and Zhao, 2015]. So all the above-mentioned studies show, at least, some persistence in atmospheric temperature. It is interesting to note that for islands, the exponent shows a broader distribution, varying from 0.65 to 0.85, with an average value close to 0.8 [Eichner et al., 2003]. Some studies report scaling exponents up to ≈1 for oceans [Bunde and Havlin, 2002; Fraedrich and Blender, 2003; Zhang and Zhao, 2015].

Varotsos, Efstathiou, and Cracknell [2013] studied the monthly means of the hemispheric temperature anomalies of land surface air 1880-2011 measured at heights of 1.25–2 m at meteorological stations [Hansen et al., 2010]. They obtained scaling exponents 0.75 for northern hemisphere (NH), 0.73 for southern hemisphere (SH) and 0.80 for global anomaly. They also studied zonal temperature anomalies, but used yearly averaged data, which consists only 131 values. Their analysis gave scaling exponents between 0.51-0.79. According to their studies, the exponents were gradually increasing as a function of latitude and more pronounced at the southern hemisphere. As we show later, our results are opposite such that scaling exponent decreases poleward and somewhat similarly at both hemispheres. This is because we use monthly averaged data, and consequently longer time series with finer resolution. According to many studies, the analyzed time series must be long enough to get reliable results, because the scaling is determined as an asymptotic behavior [Caccia et al., 1997; Weron, 2002; Rea et al., 2009; Grech and Mazur, 2013].

Shao and Ditlevsen [2016] studied the scaling of glacial and interglacial temperature time series, and found that glacial time series shows multifractal structure with generalized Hurst exponent $\alpha_q$ changing with the moment q. For example, they found $\alpha_{-2} \approx 1.4$ and $\alpha_2 \approx 1.2$. On the other hand, the behavior of Holocene temperature time series is monofractal with scaling exponent $\alpha \approx 0.7$, in line with other studies mentioned earlier. They, however, studied only local monthly record from Oxford and Prague as last 150-year temperature measurement. Furthermore, the Holocene record they used was less than 600 points.

Recently Blesić, Zanchettin, and Rubino [2019] made a comprehensive analysis of near-surface air temperature anomalies of Met Office HadCRUT4 and the NASA GISS Land–Ocean Temperature Index (LOTI) datasets and found that, excluding polar and parts of sub-polar regions (because of their substantial data inhomogeneity), the global temperature pattern is long-range autocorrelated. In their analysis, they found that scaling exponent decreases when going from equatorial towards higher latitudes. They confirmed the existence of a land–ocean contrast in persistence [Bunde and Havlin, 2002; Fraedrich and Blender, 2003] with marine data showing an appreciably more pronounced long-range persistence than land data. Their analysis is based only on second order polynomial detrended fluctuation analysis (DFA2).

There are fewer studies about the persistence of precipitation, and controversy about the (non)existence of the short or long-term correlation (persistence) in the precipitation data [Fraedrich and C., 1993; Pelletier and Turcotte, 1997; Fraedrich and Blender, 2003; Kantelhardt et al., 2006; Bunde et al., 2013; Ault et al., 2013]. This is because precipitation is part of a very complex circulation of dry and moist air masses. Another reason for the controversy is that the time resolution and length of the data vary in different analyses. At least it seems that the persistence of the precipitation is scale dependent. For example, Yang and Fu [2019] studied hourly-based precipitation and found that there is a crossover at the timescale of 200 hours such that scaling exponent is about 0.74 below this timescale and about 0.54 above this timescale. Markonis and Koutsoyiannis [2016] examined all available precipitation records over Europe with length above 200 years, as well as the CRU gridded data and found a mean for $\alpha$ coefficient close to 0.6, suggesting very weak long-term persistence for their annually averaged data.

In this article, we study temperature and precipitation anomalies extending from local to continental, hemispheric and global averages. Using detrended fluctuation analysis (DFA), we find that the scaling exponent increases when going from local records to averages of the wider areas. As far as we know this is a new result not reported in the earlier literature. Especially, the analyses of precipitation consist new insights.

We use monthly averaged data in all scaling analyses and for the years 1910-2019 if not otherwise stated. This paper is organized as follows. Section 2 presents the data and methods used in this study. In Section 3, we study persistence of the continental temperature and precipitation and in Section 4 the regional temperature and precipitation anomalies of the United States. In Section 5 we show that the wider the analyzed area, the more persistent the temperatures and precipitations. In Section 6 we present zonal and hemispheric analyzes of persistence in temperature and precipitation, and give our conclusions in Section 7.

# 2. Data and methods

## 2.1 Data

We use data from NOAA (National Oceanic and Atmospheric Administration) web site for US local and regional temperature and precipitation analysis (for regions, see Figure 1) and for hemispheric and continental temperature analysis (NOAA_2020), and for land-ocean temperature analysis NASA GISS Land–Ocean Temperature Index (LOTI) data. For continental precipitation analysis, we use data from University of East Anglia Climatic Research Unit: CRU TS4.01 (Harris and Jones, 2017). For zonal scaling analysis (temperature and precipitation), we use monthly averaged data for 1900-2017 (Willmott and Matsuura, 2018, Terrestrial Air Temperature and Terrestrial Precipitation for land area: 1900-2017 Gridded Monthly Time Series V 5.01, added 6/1/18). In zonal analysis, both hemispheres are divided into four latitude zones 0-24, 24-44, 44-64 and 64-90 degrees. The local data for Australia, UK, Czech and Finland are retrieved from http://www.bom.gov.au/climate/, https://www.metoffice.gov.uk/research/climate/, https://www.chmi.cz/historicka-data/pocasi/praha-klementinum? and https://www.ecad.eu, respectively. For comparison to land measurements, we use Gistemp Land-Ocean Temperature Index [LOTI, Lenssen et al., 2019].

Monthly temperature (in C$^{o}$) and precipitation (mm) anomalies are calculated by subtracting the average value of the corresponding month of the base period from the monthly temperature values. We normalize the data such that the base period is the whole interval, i.e., the mean value of the anomalies is zero. We do this to confirm that there is no bias outside the base period [Sippel et al., 2015; Lenton et al., 2017]. There are also other methods to deseasonalise the data [Varotsos, Assimakopoulos, and Efstathiou, 2007], but the afore-mentioned method is the most common.

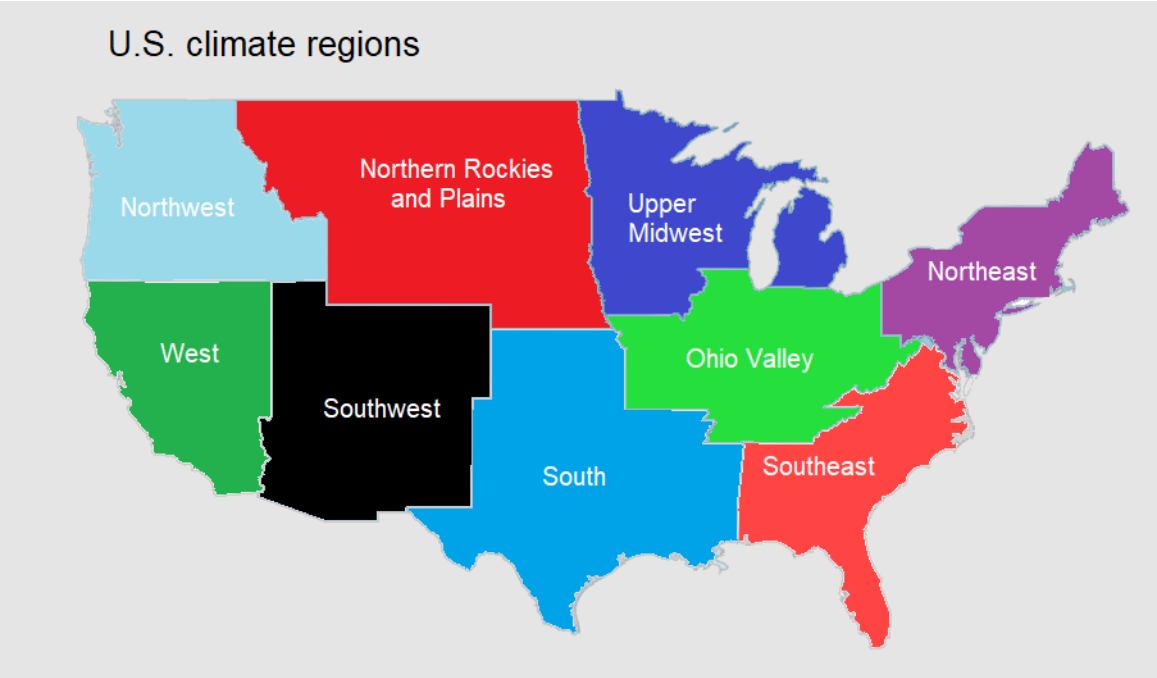

**Figure 1.** Climate regions of the USA (from NOAA National Centers for Environmental information, Climate at a Glance: Regional Mapping, published January 2019, retrieved on February 5, 2019 from https://www.ncdc.noaa.gov/cag/).

## 2.2 Rescaled range and detrended fluctuation analysis

The rescaled range (R/S) analysis was first developed to study the fluctuation of water reservoirs [Hurst, 1951; Feder, 1988], but it soon became a favorite method as a measure of long-term memory of time series. Let us have a time series y (i), i=1, 2,..., N. For each time lag τ (1≤ τ ≤N), we can define the cumulative departure from the mean

$$X(\tau) = \sum_{i=1}^{\tau} (y(i) - \langle y \rangle_\tau) \quad , \tag{1}$$

where

$$\langle y \rangle_\tau = \frac{1}{\tau} \sum_{i=1}^{\tau} y(i) \tag{2}$$

We define the range of the cumulative deviation from the mean as

$$R(\tau) = \max_{i\in[1,\tau]} X(i) - \min_{i\in[1,\tau]} X(i) \quad , \tag{3}$$

*R* presents the maximum deviation inside the lag time τ. We then calculate the standard deviation in this time interval

$$S(\tau) = \left(\frac{1}{\tau}\sum_{i=1}^{\tau}[y(i) - \langle y\rangle_\tau]^2\right)^{½} \quad , \tag{4}$$

The rescaled range is defined as the ratio R/S, which should scale asymptotically with the lag τ by the power law

$$R/_S \propto \tau^\alpha \quad , \tag{5}$$

where α is called the Hurst exponent [Taqqu, Teverovsky, and Willinger, 1995; Valverde, Castellanos, and Quintanilla, 2003]. Hurst exponent α = 0.5 describes increments of the random walk process (known also as Brownian motion) with no correlation between past and future. If the exponent α is greater than 0.5, the system is persistent, and the closer the exponent is to one the longer is the memory of the system. If the exponent α turns to be smaller than 0.5 the system is said to be anti-persistent, i.e. high values tend to follow small values and vice versa.

When applying the R/S analysis the time series needs to be stationary [Kantelhardt et al., 2002; Graves et al. 2017; Garcin, 2017]. Time series of the temperature anomalies are, however, not strictly stationary; the variances of the anomalies do not change much, but the means increase, of course, mainly because of global warming. That is why the temperature anomaly data do not fulfill the demand for strong stationarity, but they are trend-stationary. For this reason, we use a refined method to calculate Hurst exponents by the detrended fluctuation analysis (DFA), which is suitable also for non-stationary time series. For time series $y(t_i)$, $i = 1,\ldots,N$ let $\bar{y}$ be the mean value of the whole interval. We calculate the cumulative sum, i.e., sum of the deviations from the mean

$$Y(t_k) = \sum_{i=1}^{k}(y(t_i) - (\bar{y})) \quad . \tag{6}$$

Next we divide $Y(t_k)$ into non-overlapping segments each of length L. Elements of $Y(t_k)$ are renamed as $Y_{j,l}(t_k)$, where $j = 1,2,\ldots,L$ and $l = 1,2,\ldots,m$, where *m* is the number of segments. Hence $k = (l-1)\,L + j$ and $N = L\bullet m$. Each segment $(Y_{*,l})$ is now fitted with a polynomial of order *n*, i.e., $(P^n_{*,l}\,(t_k))$. An average fluctuation measure for all segments of length *L* is determined as

$$F(n, L) = \left[\frac{1}{mL}\sum_{l=1}^{m}\sum_{j=1}^{L}(Y_{j,l} - P^n_{j,l})^2\right]^{½} . \tag{7}$$

After calculating the fluctuation measure *F(n,L)* for different values of *L* it should scale as with the length *L* by the power law

$$F(n, L) \propto L^\alpha \quad , \tag{8}$$

for large values of *L*. The global scaling exponent α is here calculated with a robust regression using bi-squared weight function [Habib et al., 2017]. The confidence interval is calculated as

$$\pm t \cdot {}^{s}/_{\sqrt{N}} \quad , \tag{9}$$

where *t* is *t*-test value with confidence level 95% and corresponding degrees of freedom, *s* denotes standard deviation, and *N* the number of points in the regression analysis. In DFA the scaling exponent is conventionally marked by α, but it is the same as H (after Hurst) used traditionally in R/S analysis.

The problem with experimental data is the asymptotic nature of scaling. As stated earlier this is why we use always (if possible) time series with more than 1000 data points, which is long enough for reliable analysis [Caccia et al., 1997; Weron, 2002; Rea et al., 2009; Grech and Mazur, 2013]. Therefore, we use monthly average data to get enough data points for the scaling analysis.

Figure 2a shows $F(n,L)$ for the temperature anomalies 1895-2017 of Upper Midwest region of the USA as a function of scale ($L$) and $n = 1,..,6$ (called DFAn). Notice that for small scales there is deviation from the linear dependence, i.e., $F(n,L)$ is too small. This problem tends to over-estimate the scaling exponent, especially for short time series. Notice that in this case DFA6 behavior is strange at small scales. To correct this behavior we use the modification suggested by Kantelhardt et al. [2001]. The correction term is defined by

$$K_{\alpha}^{n} = \frac{\langle [F(n,L)]^2 \rangle L'^{\alpha}}{\langle [F(n,L')]^2 \rangle L^{\alpha}} \tag{10}$$

for $L' \approx N/20$ with $N$ the length of the time series. Here $\langle \ldots \rangle$ denotes the average of different sequences. It turns out that correction term depends only weakly about α. Therefore, we can use uncorrelated data for determining the correction term, $K_{½}^{n}(L)$. This is obtained most easily by analyzing the randomly shuffled version of the corresponding time series, because the long-range correlations are destroyed in this operation. The modified version of the fluctuation measure is now

$$F_{mod}\,(n,L) = \frac{F(n,L)}{K_{½}^{n}(L)} \tag{11}$$

Figure 2b shows the modified fluctuations $F_{mod}\,(n,L)$ of the Upper Midwest region anomalies 1895-2017 as a function of scale $L$. The changes in the fluctuations are mostly seen in the small scales and higher polynomials.

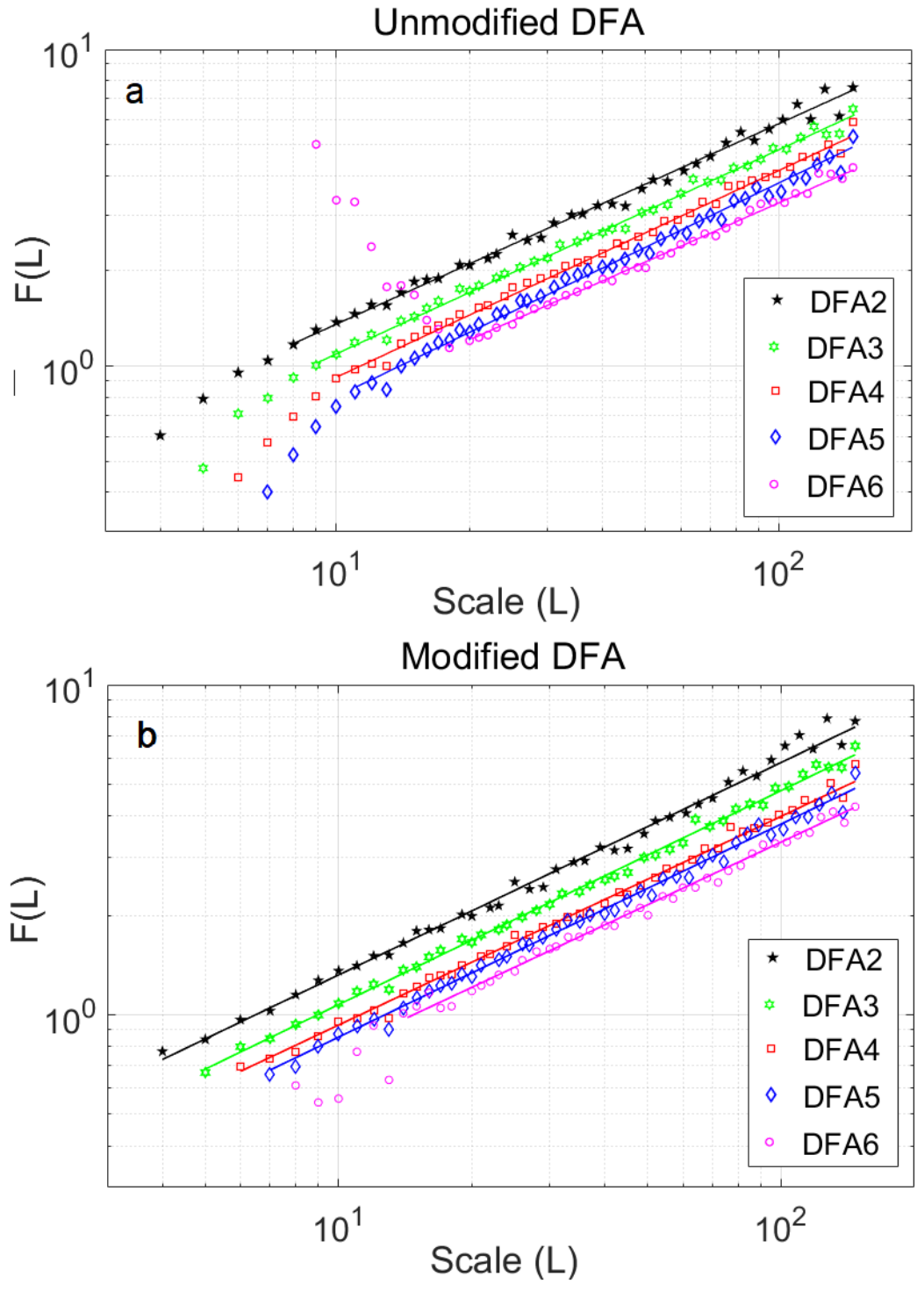

**Figure 2.** a) Average variation $F(n, L)$ as a function of scale ($L$) for the temperature anomalies 1895-2017 of Upper Midwest region of the USA for regular (unmodified) DFAn analyzes. b) Same as figure a), but calculated with modified DFAn.

Figure 3 shows how the polynomial fitting works for the cumulative sum of North America temperature anomaly during 1910-2017. Figure 3a shows the cumulative sum for the whole interval (blue curve) with ten fitted second order polynomials (black curves) using length L = 130 (months). The detrended anomaly, i.e., the residual of cumulative sum minus polynomial $Y_{j,k} - P^2_{j,k}$ , $j = 1, 2,..., L$ and $k = 1,2,..,10$ is shown as red curve in figure 3a and as magnified (red time series) and compared to the original anomaly (blue time series) in figure 3b. In this paper we use n = 2, 3, 4, 5 in DFAn analysis, because first order polynomials seem to under-fit, at least, in the longer segments (it finds only linear trends), and sixth order has difficulties in short segments (overfitting) probably due to Runge-effect. Runge phenomenon occurs as an oscillation at the edges when fitting equispaced time series with higher polynomials [Runge, 1901]. This effect can be minimized using denser set of points at the edges. We, however, use in the next analyses DFAn with n = 2, 3, 4 and 5, which seem to work well as seen from the Figure 2b.

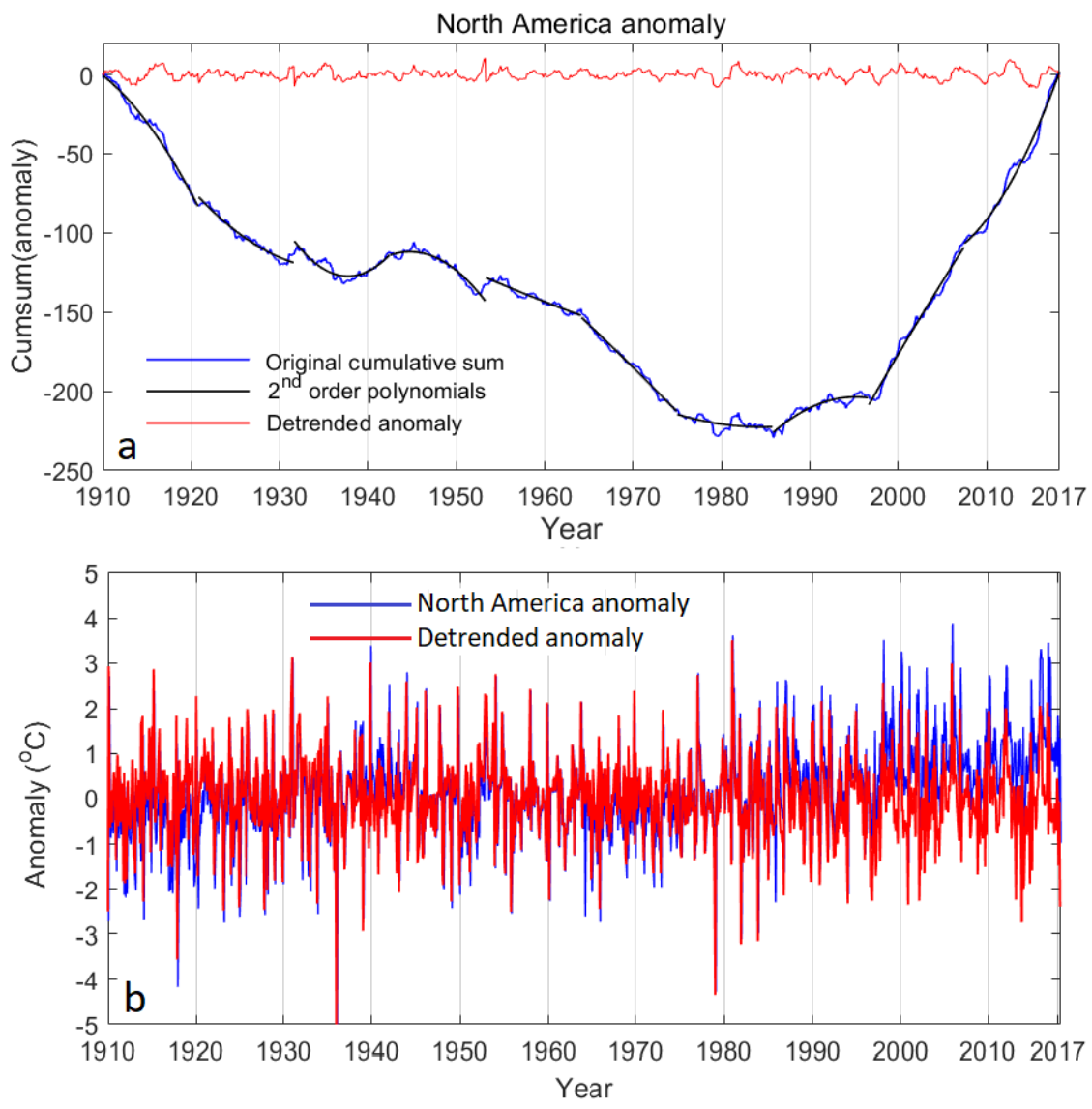

**Figure 3.** a) Profile of North America (NA) temperature anomalies with piecewise second order polynomial fittings. b) Original NA temperature anomaly (blue) and detrended temperature anomaly (red).

## 2.3 Fractional Gaussian noise

Before starting the analyses of natural data we analyzed the used DFA method for generated fractional Gaussian noise (FGN) with different values of H. We calculate fractional Brownian motion from the autocovariance function

$$C(k) = \frac{\sigma^2}{2} (|k+1|^{2\alpha} - 2|k|^{2\alpha} + |k-1|^{2\alpha}), k = 1, 2, 3,... \quad (12)$$

through circulant matrix embedding method [Caccia et al., 1997; Kroese and Botev, 2015]. Figure 4 shows measurements of scaling exponents (α) for of ten realizations of FGNs with known α. The average values for measured scaling exponent are nearest to the theoretical value for α = 0.5 and for α = 0.9, and somewhat smaller than theoretical value for α = 0.6, 0.7 and 0.8. In the table 1, we show the average values and the root mean square deviations (RMSD), i.e., the difference between measured values and predicted value for all values of α. While the used method is known to be robust, it is evident that the variation in the values of α is mainly due to inaccuracy of the DFA method.

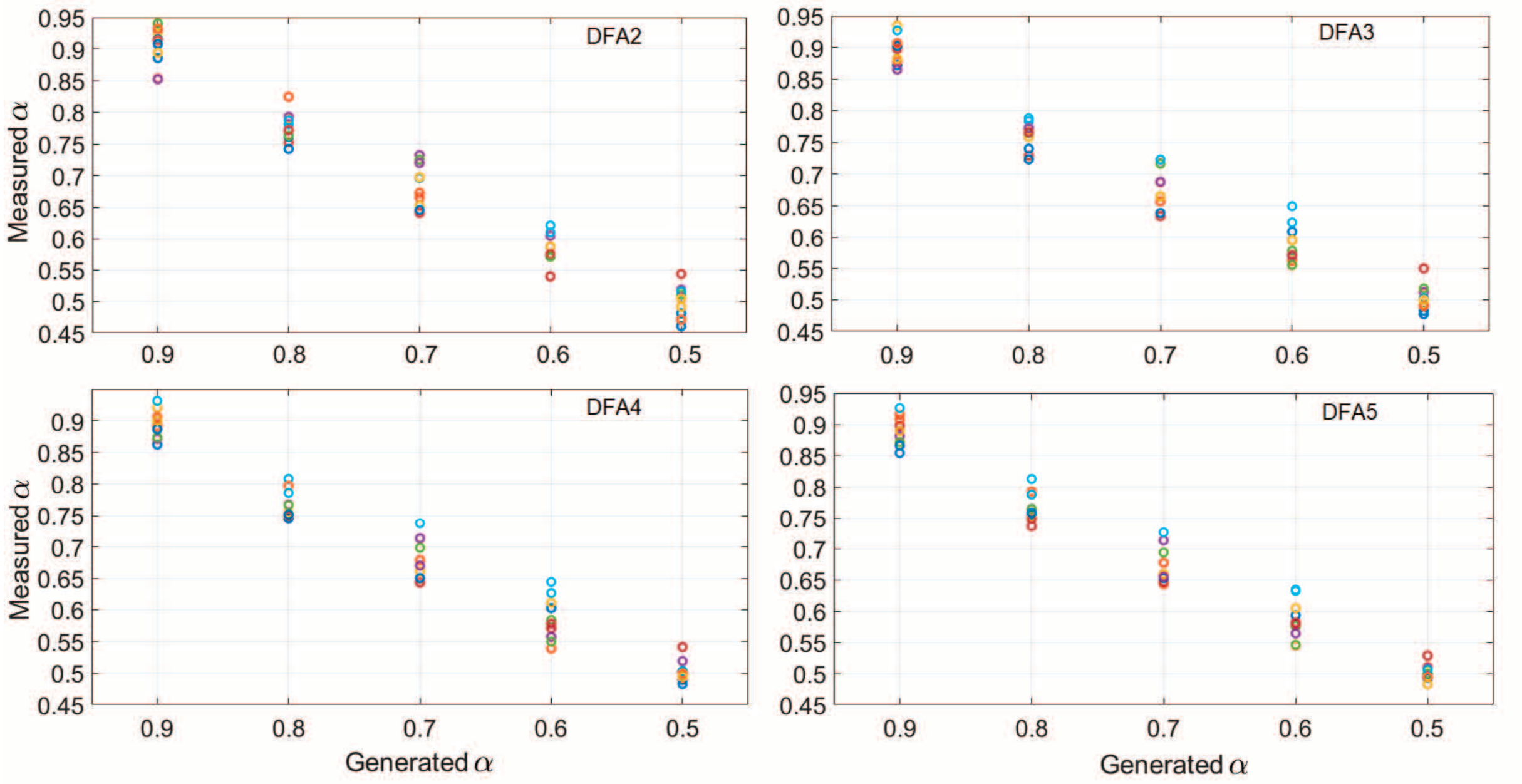

**Figure 4.** DFAn analyzes with n = 2, 3, 4 and 5 for ten realizations using H = 0.5, 0.6, 0.7, 0.8 and 0.9.

| Gen α / Meas α | 0.5 | 0.6 | 0.7 | 0.8 | 0.9 |
|---|---|---|---|---|---|
| DFA2 | 0.500 | 0.588 | 0.685 | 0.777 | 0.902 |
| RMSD | 0.029 | 0.026 | 0.036 | 0.032 | 0.029 |
| DFA3 | 0.504 | 0.587 | 0.679 | 0.762 | 0.896 |
| RMSD | 0.029 | 0.033 | 0.037 | 0.045 | 0.022 |
| DFA4 | 0.502 | 0.586 | 0.677 | 0.767 | 0.894 |
| RMSD | 0.016 | 0.035 | 0.037 | 0.039 | 0.022 |
| DFA5 | 0.503 | 0.586 | 0.672 | 0.767 | 0.888 |
| RMSD | 0.012 | 0.033 | 0.039 | 0.040 | 0.025 |

**Table 1.** Values of measured scaling exponents and their RMSDs for generated Gaussian noises with α = 0.5, 0.6, 0.7, 0.8 and 0.9.

## 3. Temperature and precipitation anomaly analysis of continents

In order to study the accuracy of the used method, modified DFA, we calculate ten estimates for the continental temperature and precipitation scaling exponents. Figure 5 shows the scaling exponents for ten estimates of the continental temperature anomalies of 1910-2016 (blue stars), and precipitation anomalies (black circles) of 1910-2016, and 95% confidence bars as red lines. Note the magnification for Europe DFA3 only to show that the points are clearly inside the limits. We show here only DFA3 and DFA4 in Figure 5a and b, respectively, but calculate the average values using also DFA2 and DFA5. The most striking feature is the high value of for Africa. The average of all 40 estimates is α = 0.862 with standard deviation 0.015, which is well inside the error limits calculated from formula 9, which are of the order of 0.02-0.03. (We use here standard deviation instead of RMSD, because we do not know the predicted value a priori). The high value is caused by the situation of Africa within 30 degree on both sides of equator. It is understandable that this leads to smaller variation of the anomalies compared to other continents, and consequently more persistent behavior of the temperatures. The scaling exponents of SH continents are 0.723 (std = 0.0079) for Oceania and 0.766 (0.026) for South America. Note that southern hemisphere scaling exponents are somewhat higher than scaling exponents of NH continents, i. e. Asia, $\alpha = 0.691$ (std = 0.032), Europe, 0.701 (0.0095) and North America, 0.637 (0.011). We can say that Africa has high persistence and the other southern hemisphere continents moderate persistence (long-term correlation) in their temperatures. On the other hand, the northern hemisphere continents are only somewhat persistent. The relatively higher standard deviations of South America and Asia are due to different results for different order polynomials (n) in DFAn. Note for example, that the scaling exponents of DFA3 and DFA4 for Europe are almost the same but lower than 0.7 and higher than 0.7 for Asia, respectively.

The scaling exponents of the precipitations are much smaller than the scaling exponents of the temperatures, except for North America. The average values for the scaling exponents of four DFAs are Africa, 0.600 (std = 0.009); Asia, 0.598 (0.012); Europe, 0.547 (0.0038); North America 0.675 (0.0087); Oceania 0.667 (0.028) and South America 0.624 (0.017). The reason for the high standard deviations for southern hemisphere continents is again the same as stated earlier for SA temperature scaling exponent. Interestingly, there is a slight persistence in precipitation for all other continents than Europe. In order to see if the persistence is real and not due to estimator itself, we randomly shuffled the anomalies of the temperature and precipitation. Figure 5c and d show the scaling exponents of the shuffled anomalies of temperature (blue stars) and precipitation (black circles) for DFA3 and DFA4, respectively. The scaling exponent are near the value 0.5 as expected for the random time series. Note, however, that the distributions of the anomalies may differ from Gaussian distribution. Nevertheless, we use log-log linear piecewise fitting for the scaling exponent in all our analyses, and adjust the scaling regime depending on the length of the time series. For example, visually this means in the unmodified DFAn (n = 2,3,4 and 5) that the scaling region in Figure 2a could be L from 11 to 150, but for modified DFA of Figure 2b the region could be somewhat wider, i.e. L = 7-150. Note that while lower polynomials work better with short intervals L, they work worse for long L. This is especially seen for DFA2 with L >100.

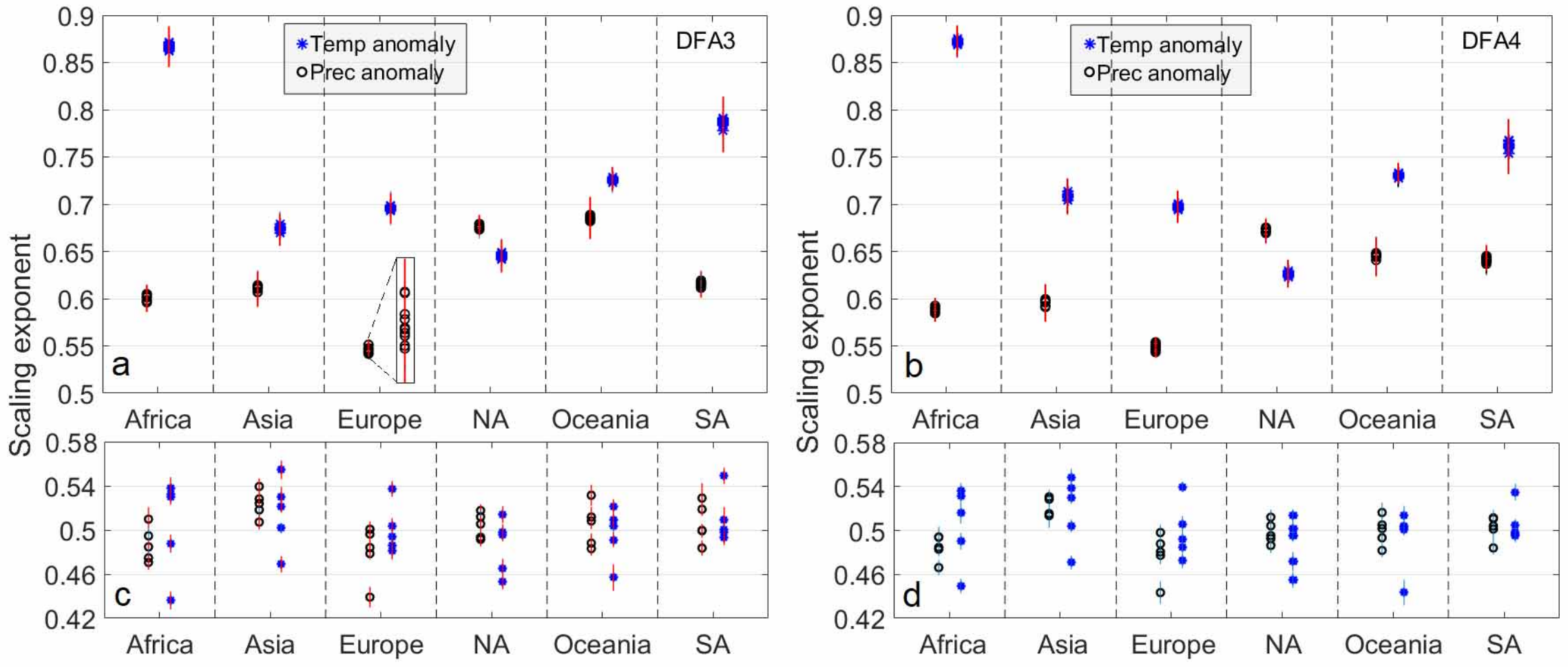

**Figure 5.** a) and b) Scaling exponents of the continental DFAn analyzes for ten measurements of precipitation and temperature anomalies using n = 3 and 4, respectively. c) and d) Five measurements for randomly shuffled time series of the same anomalies using n = 3 and 4, respectively.

# 4. USA regional temperature and precipitation anomalies

As it turned out that North America is the only continent, which has persistence similar for temperature and precipitation, we analyze next the climate regions of USA. Figure 7 shows the temperature and precipitation anomaly scaling exponents of DFA2, DFA3, DFA4 and DFA5 analyses for all nine contiguous US continental climate regions (see Figure 1, note that Alaska is missing) in 7a, b, c and d, respectively. The blue stars show the temperature anomaly scaling exponents calculated with the modified DFAn and the red vertical bars show the 95% significance of the results (calculated from formula 9). Note that the scaling exponents increase slightly with increasing fitting polynomial. It is, however, clear that OV (Ohio Valley), S (South) and SE (Southeast) have the smallest values for the temperature scaling exponents, i.e. 0.58, 0.60 and 0.58 on the average, respectively. These values are quite near the exponent for random noise 0.5. The other regions have scaling exponents: NE (Northeast) 0.63, NRaP (Northern Rockies and Plains) 0.61, NW (Northwest) 0.66, SW (Southwest) 0.66, UM (Upper Midwest) 0.64, and W (West) 0.66 on the average. These regions show weak persistence in the temperature during 1910-2019. Note that the westernmost climate regions, NW, W and SW, have highest and equal scaling exponent, although they belong to diverse climate areas [Kottek et al., 2006, Peel et al., 2007], (see also Figure 6).

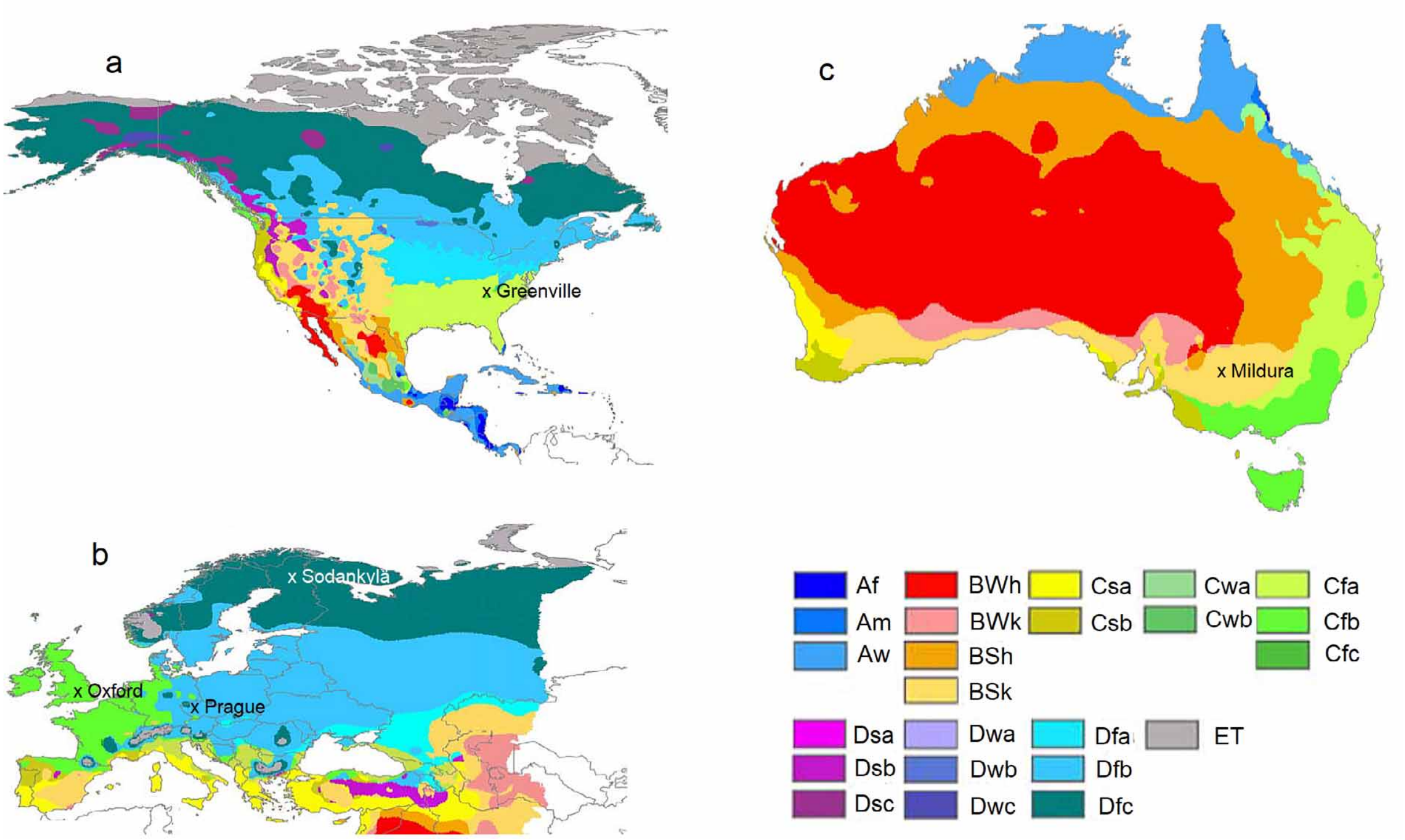

**Figure 6.** Köppen-Geiger climate types of North-America, Europe and Australia (from Peel, Finlayson, and McMahon, [2007]). The analyzed local sites are marked in the map. The symbols are explained in Table 2.

The black circles show the corresponding DFAn values for the precipitation anomalies. It is evident that the persistence of the precipitation is smaller than the persistence of temperature for other climate regions than OV, S and SE. These regions have scaling exponents almost the same size for both temperature and precipitation anomalies. Note that S and SE belong totally, and OV partly to the warm oceanic/humid subtropical and partly to humid continental climate area [Kottek et al., 2006, Peel et al., 2007, Chen and Chen, 2013]. From the other six climate regions, the northern climate regions NE, UM and NW have largest difference between persistence of temperature and precipitation. Note especially, that NE has precipitation scaling exponent smaller than 0.5. NRaP is the only region belonging mainly to cold semi-arid climate type, and has smallest difference in persistence of temperature and precipitation (except OV, S and SE). The southwestern climate regions W and SW are very near each other also in the persistence of precipitation. We may say, however, that all the other regions, except NE, are slightly persistent in monthly precipitation.

The lower panels of the figures show scaling exponents for five measurements of randomly shuffled precipitation anomalies (circles) of NE, NW, S, SW, and W, and for five measurements of randomly shuffled temperature anomalies of NRaP, OV, SE and UM (triangles). It is clear that the slight persistence, which exists, at least, in some temperatures is due to long-range correlation and not to the intrinsic property of the estimator itself.

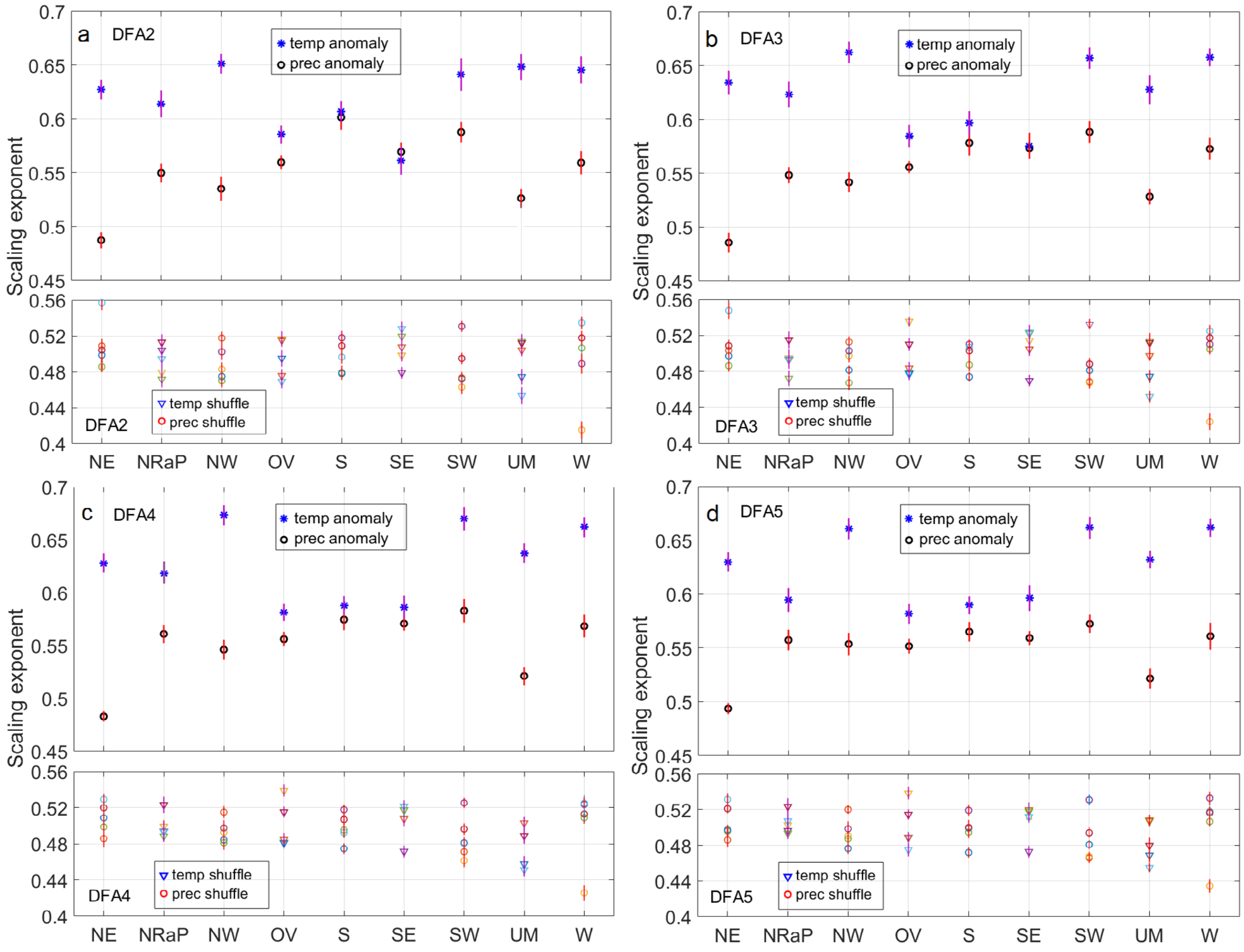

**Figure 7.** Blue stars (black circles) show the temperature (precipitation) anomaly scaling exponents of DFAn analyzes with a) n=2, b) 3, c) 4 and d) 5 for the US climate regions. The red lines show the 95% confidence limits calculated from formula. The lower panel of the figures show scaling exponents of the five realizations of randomly shuffled data for precipitation anomaly (triangles) of five regions, and temperature anomaly (squares) of four regions.

## 5. Local to global anomalies

In this section, we calculate scaling exponents starting from local area (citywide) and going through statewide and countrywide to hemispheric land temperature anomalies. Figure 8 shows the scaling exponents of the anomalies for Greenville (Gr, average $\alpha = 0.56$), South Carolina (SC, $\alpha = 0.58$), Southeast (SE, $\alpha = 0.58$), USA ($\alpha = 0.60$), and Northern hemisphere (NH, land area, $\alpha = 0.71$). The scaling exponents increase monotonically when going from local to larger areas, although the exponent of the whole USA is smaller than exponent of Southeast region for DFA4 and DFA5. This may be due to bad fitting with higher polynomials for the temperature anomaly of the USA. This is also the reason why we should calculate the scaling exponent as an average of several polynomial fittings. However, it is likely that the reason for the equal scaling exponents of Southeast and contiguous USA is the diversity of the climate regions of the USA. The precipitation anomalies show still better monotonic increase with the size of the measured area. The average scaling exponents of the precipitation anomalies are for Greenville ($\alpha = 0.52$), South Carolina ($\alpha = 0.55$), Southeast ($\alpha = 0.57$), USA ($\alpha = 0.60$) and Northern hemisphere land area ($\alpha = 0.67$). It is evident that the temperature and precipitation of Greenville, and perhaps precipitation of South Carolina, have no correlation. Note also that the error bars of Greenville are very wide. Southeast and contiguous USA are slightly persistent in precipitation and temperature. Note that Southeast and USA have scaling exponents of temperature and precipitation very near each other, i.e. precipitation exponent is higher than exponent of temperature for DFA2, about equal for DFA3, but vice versa for DFA4, and DFA5. This situation is still clearer for the whole North America, with precipitation being more persistent than temperature, as shown in the continental scaling of Figure 5.

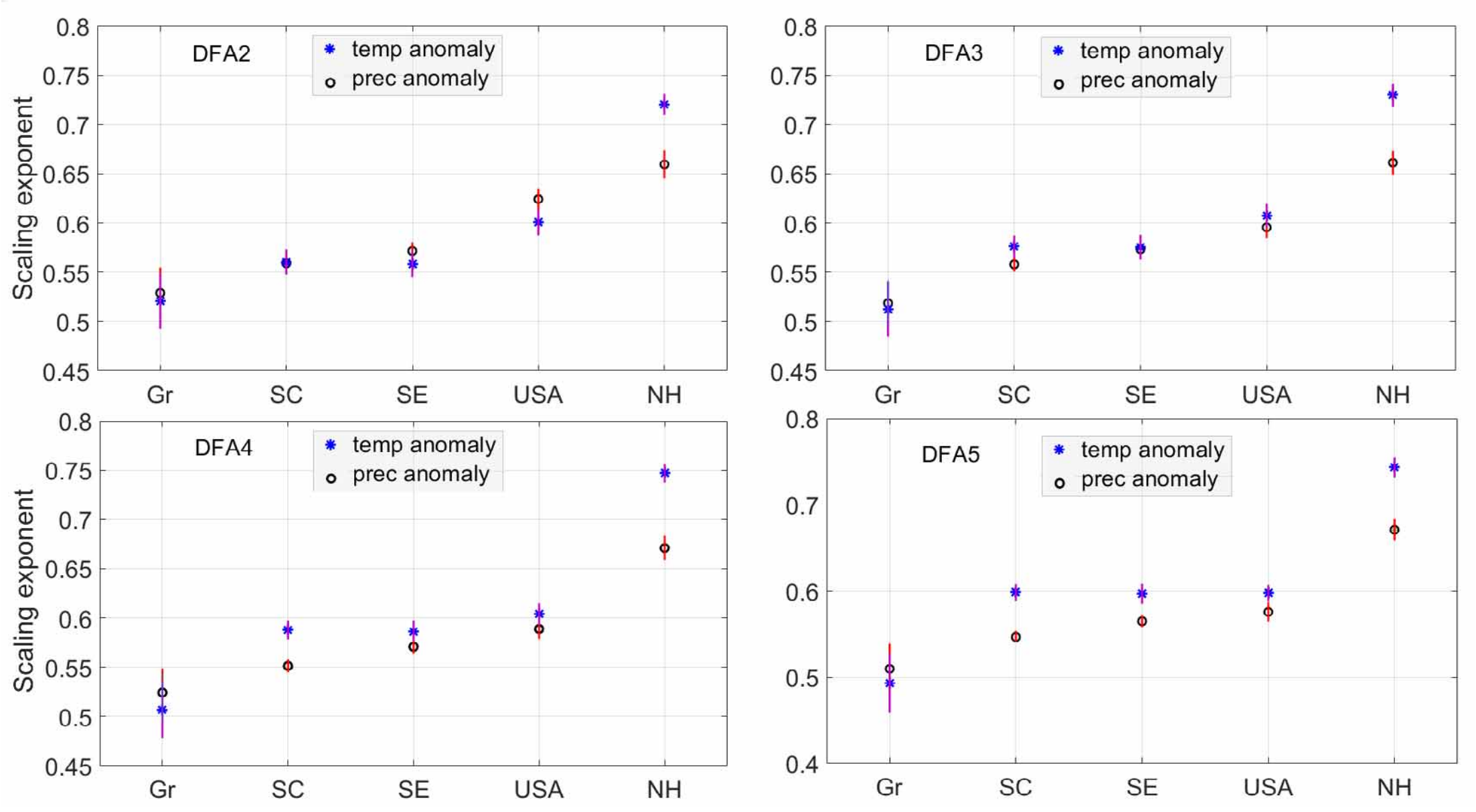

**Figure 8.** Scaling exponents for Greenville (Gr), South Carolina (SC), Southeast (SE), USA and Northern hemisphere land area (NH) of DFAn analyzes with n = 2,3,4 and 5 for temperature anomalies (blue stars) and precipitation anomalies (black circles).

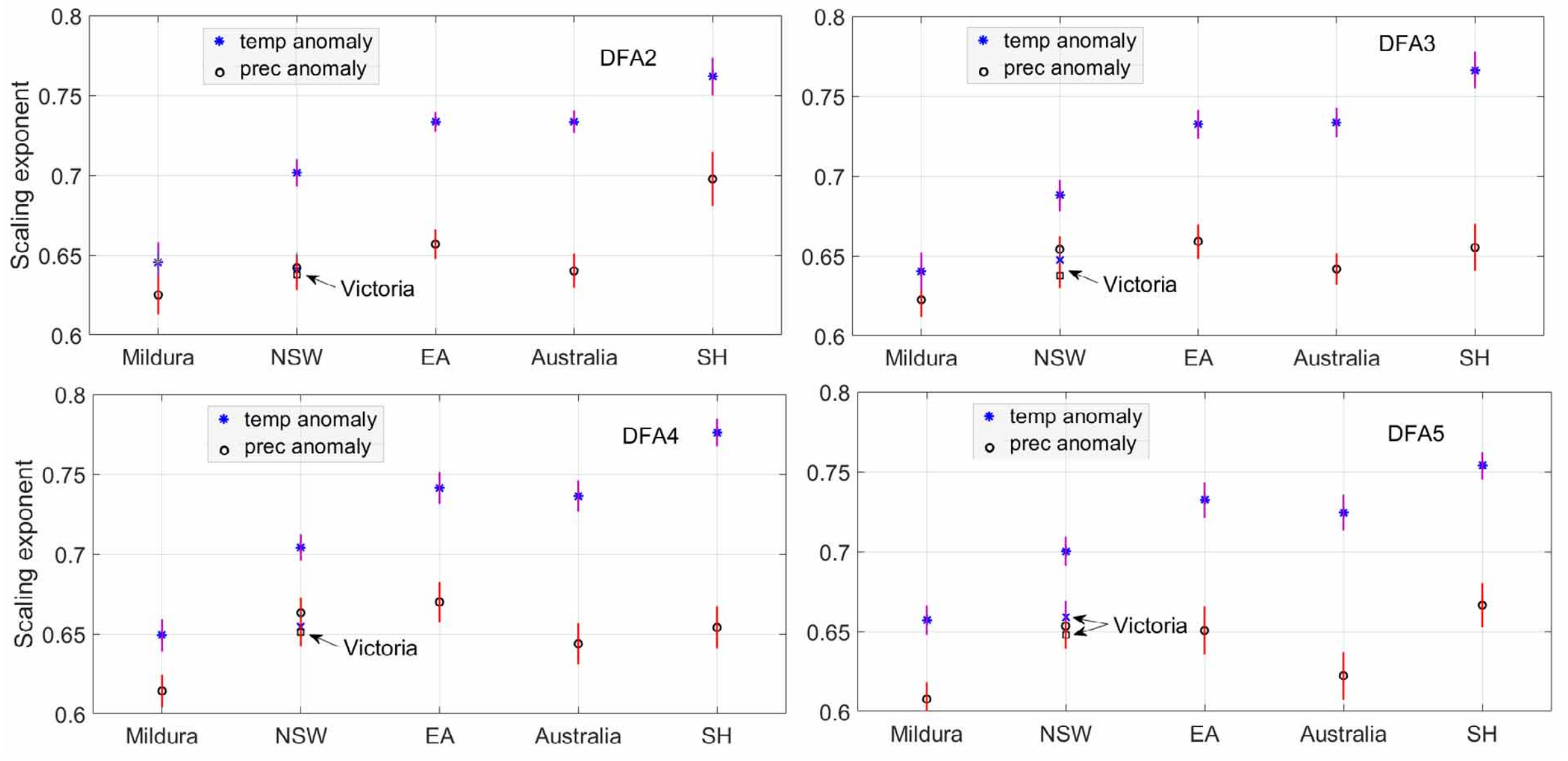

**Figure 9.** Scaling exponents for Mildura, New South Wales (NSW), Eastern Australia (EA), Australia and Southern hemisphere land area (NH) of DFAn analyzes with n=2,3,4 and 5 for temperature anomalies (blue stars) and precipitation anomalies (black circles).

In order to test if similar increase in persistence occurs in the southern hemisphere we have calculated scaling exponents for eastern Australia regions. The local site is Mildura, where measurement for temperature and precipitation measurement are available at least from 1910. Figure 9 shows the scaling exponents of temperature (blue stars) and precipitation (black circles) anomalies for Mildura, New South West (NSW), Eastern Australia (EA), the whole Australia and Southern hemisphere land area (SH). Because Mildura is located near the border of NSW and Victoria, we also show scaling exponent of Victoria (temperature as a cross and precipitation as a square). It seems that scaling exponent increases (on the average) as a function of area, except between the Eastern Australia (Queensland, NSW and Victoria) and the whole Australia. This is probably due to the heterogeneity of the Australian climate regions with deserts covering largest part of the country. The dominant climate type by land area is Arid B (77.8%), followed by temperate C (13.9%) and tropical A (8.3%) [Peel, Finlayson, and McMahon, 2013], (see also Figure 6). The average scaling exponents for temperature anomalies are: Mildura, 0.65, NSW, 0.70, Eastern Australia, 0.73, Australia, 0.73, Southern hemisphere, 0.76. The corresponding values for precipitation anomalies are Mildura, 0.61; NSW, 0.65; EA, 0.66; Australia, 0.64; SH, 0.67. Note that both temperature and precipitation anomaly scaling exponents for Victoria are at the same level as the scaling exponent of precipitation for NSW. This is interesting, because both states have similar mixture of climate types. Victoria is located, however, south, i.e. poleward from NSW, and has thus lower scaling exponent for temperature than NSW. We tested this hypothesis by analyzing the latitudes S35-S30 (latitudes of NSW), and latitudes S40-S35 (latitudes of Victoria) for the whole Australia and it showed scaling exponents about 0.71 and 0.67, respectively (not shown here). The exponents are slightly higher than those for NSW and Victoria alone, because the area is wider, i.e. constitute longitudes E115-E155. Note that NSW and Victoria have quite similar scaling exponent for precipitation, showing that precipitation is less dependent on the latitude. Mildura itself is located at dry semi-arid climate region (see Figure 6, where we show approximate locations of the analyzed local sites and Table 2). We, however, see that there is a weak persistence in precipitations and weak to moderate persistence in temperatures for different areas of Australia.

| Major group | Sub-types |
|---|---|
| A: Tropical | Tropical rain forest: Af |
| | Tropical monsoon: Am |
| | Tropical wet and dry savanna: Aw |
| B: Dry | Desert (Arid): BWh, BWk |
| | Steppe (Semi-arid):Bsh, Bsk |
| C: Mild temperate | Mediterranean: Csa, Csb, Csc |
| | Humid subtropical: Cfa, Cfb |
| | Oceanic: Cfb, Cfc, Cwb, Cwc |
| D: Continental (Snow) | Humid: Dfa, Dwa, Dfb, Dwb, Dsa, Dsb |
| | Subarctic: Dfc, Dwc, Dfd, Dwd, Dsc, Dsd |
| E: Polar | Tundra: ET |
| | Ice cap: EF |

**Table 2.** Main Köppen-Geiger climate types and their sub-types [after Chen and Chen, 2013].

From the third continent, Europe, we use three examples, Oxford in United Kingdom, Prague in Czech Republic, and Sodankylä in Finland. European countries are small (in area) compared to USA and Australia, and usually belong to only one climate sub-type. United Kingdom belongs to mild temperate oceanic climate type similar to NSW in Australia, and is near that of southeastern USA (see map in Figure 6). Czech belongs to humid continental area, with snow in winter, very similar to northeastern USA. Finland represents here the only (mainly) subarctic region, and Sodankylä is a small place in Lapland above northern polar circle.

The DFA3 and DFA4 scaling exponents of the temperature and precipitation anomalies are shown in Figure 10 for UK Oxford (UK), Prague (Czech) and Sodankylä (Finland), from top to bottom, respectively.

We see a monotonic rise in the exponents of the temperature anomalies as a function of the area size, except that Czech has slightly smaller exponent than Prague. There exists a slight persistence in temperature for these places (countries). Precipitation anomalies behave differently such that the whole country has smaller scaling exponent than local region, except for Czech. It is, however, clear that the scaling exponents of the precipitation are near 0.5 in all cases including that of the whole Europe. We may deduce that the long-term precipitation in unpredictable, except, of course in the case of seasonal variations.

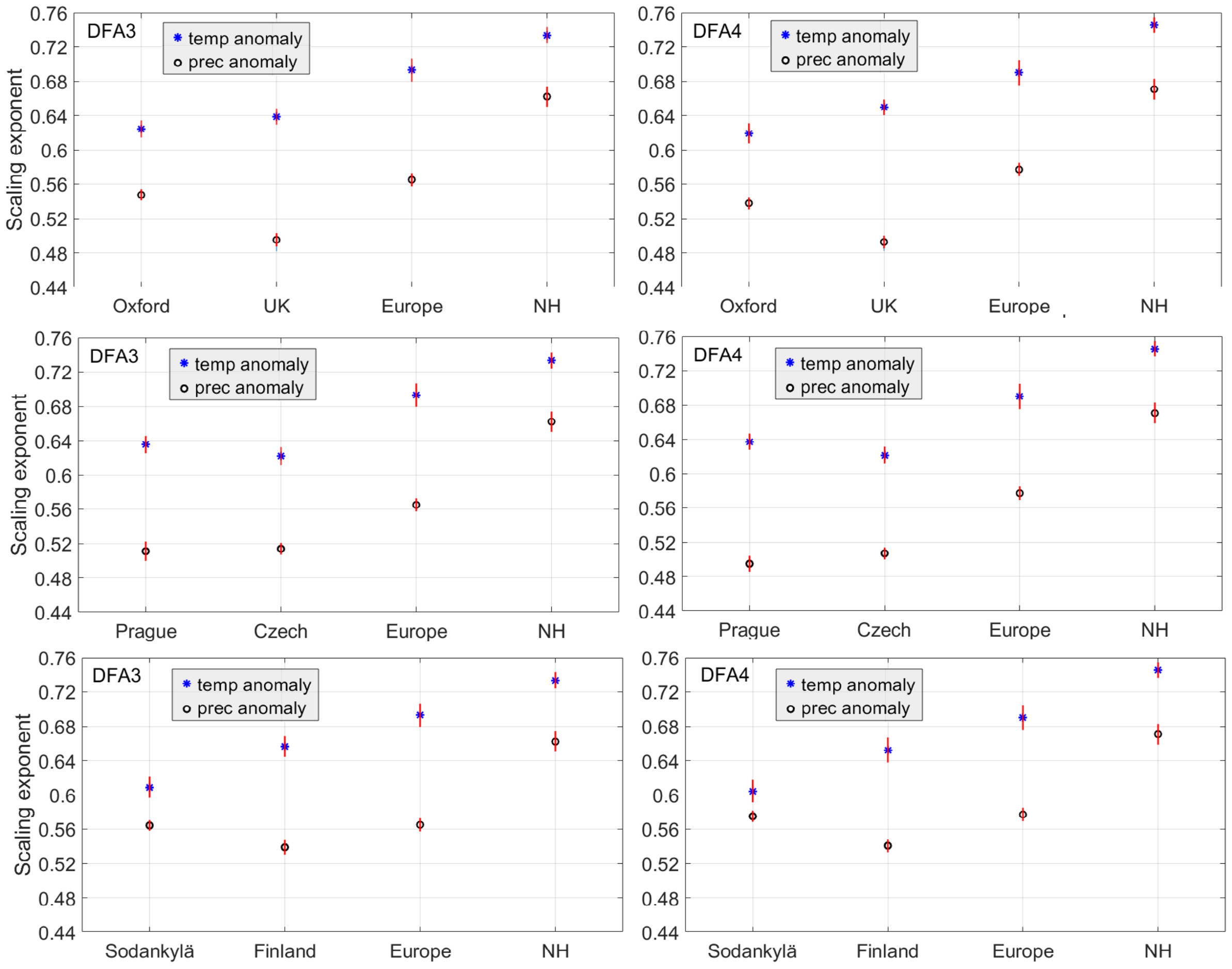

**Figure 10.** Scaling exponents for Oxford, United Kingdom (uppermost panels), Prague, Czech (middle panels) and Sodankylä, Finland (lowest panels) with Europe and Northern hemisphere land area (NH) of DFAn analyzes with n = 3 and 4 for temperature anomalies (blue stars) and precipitation anomalies (black circles).

## 6. Anomalies of the latitude zones

For the zonal scaling analysis, we use monthly averaged data for 1900-2017 [Willmott and Matsuura, 2018]. Table 3 lists the number of stations used in 2010 in temperature and precipitation measurements. The number of stations has varied from about 1500 (in 1900) to over 18000 (2010) for temperature and up to 40000 (in 1970) for precipitation, and vast majority of them are continental northern hemisphere stations.

The station temperatures were spatially interpolated to a 0.5-degree-by-0.5-degree latitude/longitude grid on land area, which consists of 85794 points, 49606 in NH and 36188 in SH (K. Matsuura, private communication). Note that, although equatorial zones had smaller amount of stations, there were lot of oceanic stations. It is clear, that because of the much larger heat capacity of water compared to air, the oceans smooth the temperature variation, and consequently cause long-range correlation in the temperatures.

Figure 11 shows the scaling exponents of temperature and precipitation for the seven latitude zones of the northern and southern hemispheres (the zone S90-S64 is omitted due to only few measurements, especially in the first half of 20$^{th}$ century). The most striking feature is the high scaling exponent on both sides of the equator, i.e., 0.95 and 1.02 on the average for temperature anomaly and 0.77 and 0.72 for precipitation anomaly of latitudes (EQ)-N24 and S24-EQ, respectively. The high values are due to equatorial climate with lot of humid rainforest areas in both sides of the zero latitude [Kottek, 2006]. Note, that they have also the widest 95 % confidence limits, at least for temperature. The scaling exponents of other zones are 0.67 and 0.61 (N64-N90), 0.70 and 0.59 (N44-N64), 0.68 and 0.55 (N24-N44), 0.71 and 0.62 (S44-S24) and 0.68 and 0.56 (S64-S44) on the average for temperature and precipitation anomalies, respectively. The reason for even higher scaling exponent around the equator, compared to scaling exponent of Africa, is that relatively many measurements are from stations surrounded by ocean. The scaling exponents of temperature anomalies decrease when going poleward from equatorial region for both northern and southern hemisphere.

| Latitude | Temperature stations | Precipitation stations |
|---|---|---|
| N64-N90 | 600 | 558 |
| N44-N64 | 6320 | 6711 |
| N24-N44 | 8846 | 13025 |
| EQ-N24 | 1073 | 990 |
| S24-EQ | 613 | 911 |
| S44-S24 | 822 | 3692 |
| S64-S44 | 42 | 34 |

**Table 3.** The amount of stations (2010) used in the temperature and precipitation zonal analyses [Willmott and Matsuura, 2018].

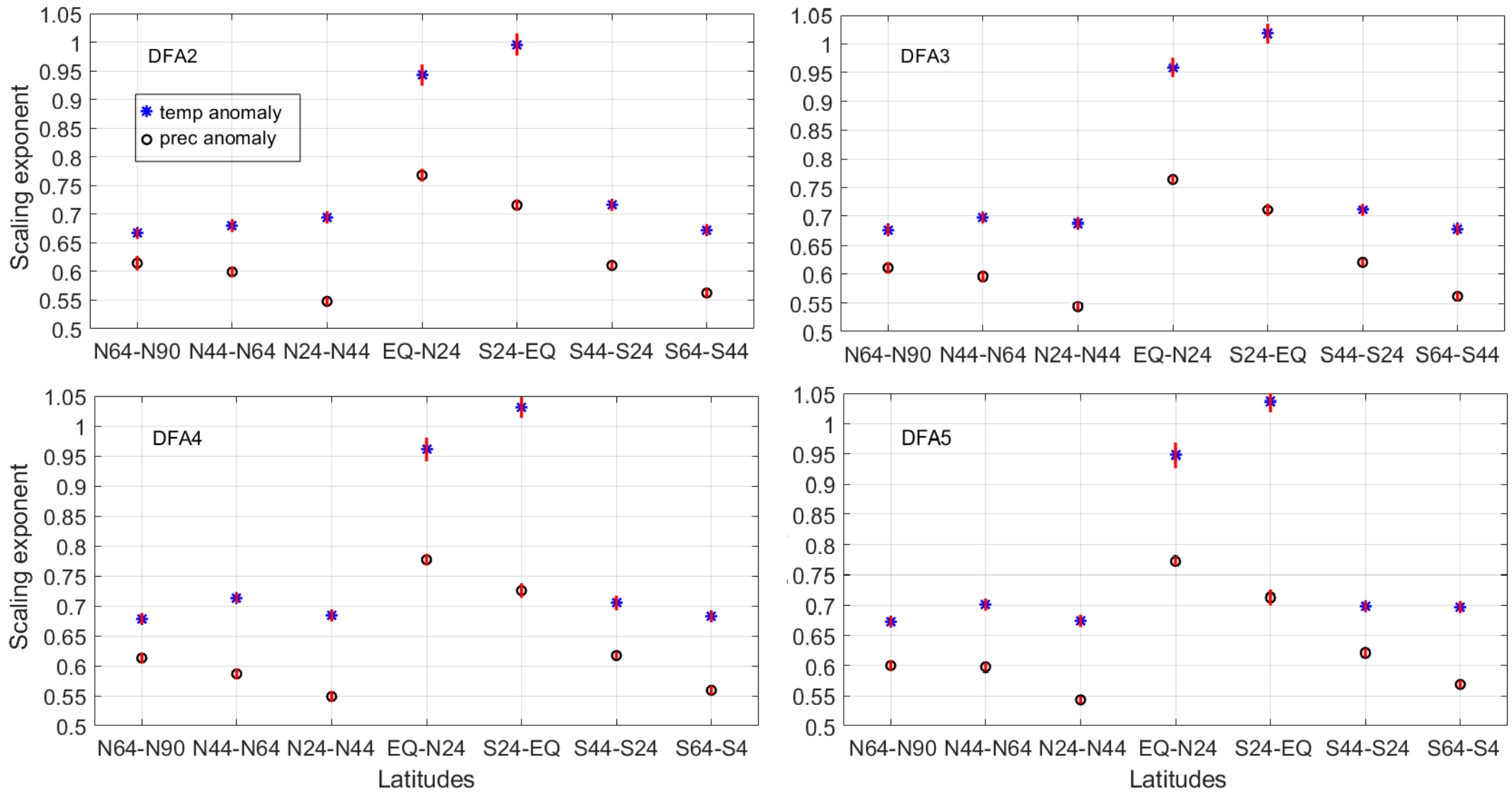

**Figure 11.** Scaling exponents for the zonal DFAn analyzes of temperature and precipitation with n = 2, 3, 4 and 5.

Note that the behavior of the scaling exponents as a function of latitude is opposite to the results by Varotsos, Efstathiou, and Cracknell [2013], who used only 131 data points of yearly averages in their study of temperature anomalies. On the other hand, our temperature results are in line with the scaling of temperature anomalies by Blesić, Zanchettin, and Rubino [2019]. Király and Jánosi [2005] obtained similar results for station wide temperature measurements of Australia. The scaling exponent of the precipitation also decreases when moving southward in the southern hemisphere. On the contrary, the scaling exponent of the precipitation anomalies behaves differently in the northern hemisphere such that it is in minimum at the zone N24-N44 and increases then towards northern pole, although the increase is small. However, while the precipitation is random in the zone N24-N44, there is a slight persistence in the zones N44-N64 and N64-N90.

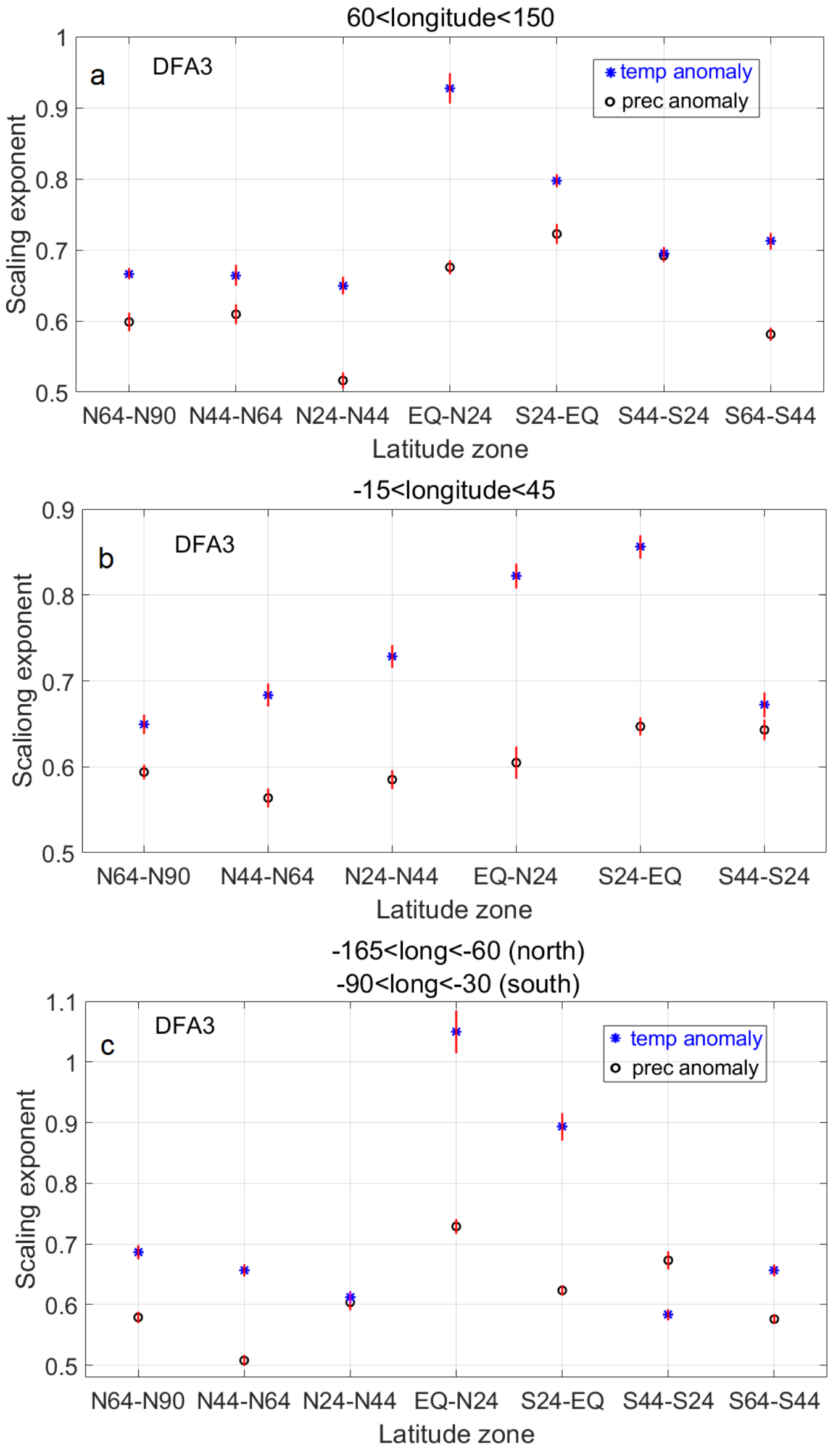

**Figure 12.** Scaling exponents for the zonal DFA3 analyzes a of temperature and precipitation a) for longitudes 60 to 150 b) for longitudes -15 to 45 and c) for longitudes -165 to -60 for northern hemisphere and -90 to -30 for southern hemisphere. (Note that panel b does not have zone S64-S44).

As the result especially for precipitation in the northern hemisphere seems so interesting we studied separately scaling exponent longitudinally at the sites of continents, i.e. longitudes 60E-160E (Asia and Australia), longitudes 15W-45E (Europe and Africa), and longitudes 165W-60W (North America) and longitudes 90W-30W (South America). Figure 12 depicts the results of these analyses. It seems that the randomness of the zone N24-N44 is due to the scaling exponent of precipitation mainly in Asia. These latitudes include very different types of climate from mountains to rainforest and deserts in Asia, i.e., the area is very heterogeneous. The largest part of this area undergoes summer monsoon, which is, of course, predictable, but this periodical phenomenon has been removed from the analyzed data.

Another zone, which is longitudinally varying is N44-N64. Its precipitation scaling exponent is at lowest, i.e. fully random (0.5), in North America. These latitudes in North America include mainly Canada and northern part of USA. They belong to continental snowy humid climate region with partly warm, partly cool summer. This seems to be combination, which leads to very random precipitation. In the temperature anomalies, the most striking detail is that only for longitudes 15W-45E the scaling exponents neatly decrease from equatorial regions poleward.

Figure 13 shows that if the ocean surface temperatures are included together with the land surface temperatures in the hemispheric and global temperature data the persistence increases considerably. The land temperatures in the Figure 13 are from NOAA (2020) database for 1910-2019 (green squares), and Willmott and Matsuura V 5.01 database for 1910-2017 (blue stars) and Land-Ocean temperatures from Gistemp Land-Ocean Temperature Index (LOTI) for 1910-2019 (magenta triangles) [Lenssen et al., 2019]. The scaling exponents for LOTI data are 0.88, 0.97 and 1.0 for northern hemisphere, southern hemisphere and globally, respectively. The corresponding land surface temperature values are 0.74 (WM) and 0.73 (NOAA), 0.88 (WM) and 0.90 (NOAA), and 0.76 (WM) and 0.79 (NOAA) for NH, SH and globally, respectively.

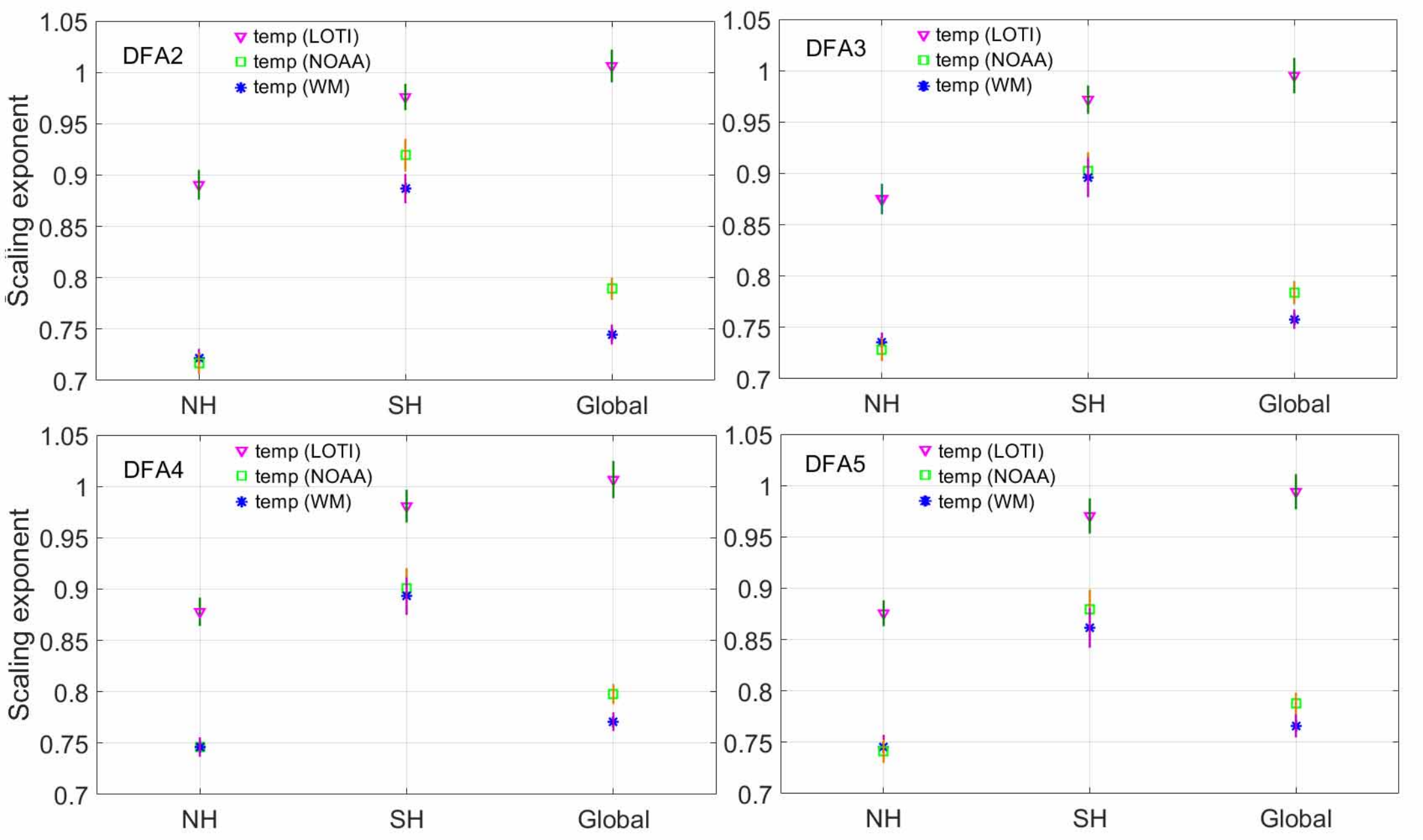

**Figure 13.** Scaling exponents of three different datasets, Land-Ocean Temperature Index for 1910-2019 (LOTI), NOAA land surface temperature anomalies for 1910-2017 (NOAA), and Willmott-Matsuura temperature anomalies of land stations (WM).

## 7. Conclusions

We have studied the long-term correlation of temperature and precipitation of the continents using detrended fluctuation analysis (DFA). We showed that the persistence (correlation) in temperature of Africa is by far higher than other continents. The scaling exponent for Africa temperature anomaly is 0.86. The other southern hemisphere continents, South America (scaling exponent, $\alpha = 0.77$) and Oceania (0.72) have also higher temperature scaling exponents than the northern hemisphere continents: Asia (0.69), Europe (0.70) and North America (0.64). We can say that Africa is strongly persistent, South America and Oceania moderately persistent and the NH continents considerably persistent in temperature. On the other hand, North America (0.68) and Oceania (0.67) have slightly higher scaling exponents of the precipitation anomalies than the other four continents: Africa (0.60), Asia (0.60), Europe (0.55) and South America (0.62). We, furthermore, have shown that the persistence is due to the long-term correlation in the data and is not an intrinsic property of the estimator itself. We note that Europe is hardly persistent in precipitation, Africa and Asia slightly persistent and North America and Oceania considerably persistent in precipitation. Note that, although the zone N44-N64 of North America is random in precipitation, North America as a whole includes also southern part of USA and Middle American countries and thus is considerably persistent in precipitation.

We showed that the scaling exponent α of the temperature anomaly increases when going from local to wider area averages. For example, the scaling exponent successively rises for Greenville, South Carolina, Southeast, USA, North America and Northern hemisphere land area anomalies. This phenomenon may be somewhat due to measurement noise being smoothed when extending the measured area. We, however, also suppose that local sudden changes are reduced when widening the area of investigation thus leading to more persistent overall behavior of the climate. This is not always so straightforward, especially for precipitation, if the climate of the larger region is heterogeneous and consists of different climate types.

We also studied the temperature and precipitation anomalies of the climate regions of the United Sates. We find an interesting difference in the behavior in temperatures for Ohio Valley, South and Southeast, i.e., the oceanic subtropical region, from other regional areas of the USA. The temperature scaling exponents of these regions are smaller than the scaling exponents for other US climate regions, showing that the temperatures of the three regions are less persistent compared to other regions. On the other hand, the persistence of the precipitation of these regions is slightly higher than the other regions.

The scaling exponents of the temperature anomalies decrease from equatorial zones poleward in both hemispheres. The exponents of the latitudinal zones are between 0.67-0.71 for S64-S44, S44-S24, N24-N44, N44-N64 and N64-N90, but the two equatorial zones S24-EQ and EQ-N24 have scaling exponents 1.02 and 0.95, respectively. These high values are due to the equatorial tropical climate type all over the world at these latitudes. The scaling exponents of precipitation anomaly are between 0.72 and 0.77 for S24-EQ and EQ-N24, respectively, but for other zones between 0.55 and 0.62. The smallest value 0.55 belongs to zone N24-N44. The diversity of the climate types in Asia leads to non-persistence in precipitation on the Asian part of the zone N24-N44. In addition, the value 0.56 for zone S64-S44 is due to the different climate types in South America in that zone. It seems that the persistence of the precipitation is more dependent on the corresponding climate type than the persistence of temperature.

We are using linear piecewise log-log fitting, which seems to work well in climate analysis. It should, however, be noted that recently, especially in biophysical applications, new methods have been developed to prevent biasing of measurement noises in evaluating anomalous diffusion exponents in subdiffusive ($\alpha < 0.5$), diffusive ($\alpha \approx 0.5$) and superdiffusive ($\alpha > 0.5$) processes. These studies consider trajectories of microscopic particles in living organs and other complex systems, and are somewhat different field of research than the present study [Jeon, Barkai, and Metzler, 2013; Kepten et al., 2015; Lanoiselée at al., 2018; Briane, Kervrann and Vimond, 2018; Bo et al., 2019; Weron et al., 2019; Janczura et al., 2020].

**Acknowledgements.** The US local, regional, hemispheric and continental temperature anomalies are retrieved 2020-09-01, from https://www.ncdc.noaa.gov/cag/. The zonal data are calculated from package: Terrestrial Air Temperature: 1900-2017 Gridded Monthly Time Series V 5.01 (http://climate.geog.udel.edu/~climate /html_pages/download.html). We use data from University of East Anglia Climatic Research Unit: CRU TS4.01 (Harris and Jones, 2017). The local climate data of Australia, United Kingdom and Czeck republic were retrieved from www.bom.gov.au/climate/, https://www.metoffice.gov.uk/research/climate/ and https://www.chmi.cz/historicka-data/pocasi/praha-klementinum?, respectively. For Sodankylä data we acknowledge the data providers in the ECA&D project. Data and metadata available at https://www.ecad.eu. The LOTI data is from GISTEMP Team, 2020: *GISS Surface Temperature Analysis (GISTEMP), version 4*. NASA Goddard Institute for Space Studies. Dataset accessed 2020-12-03 at https://data.giss.nasa.gov/gistemp/. We are grateful to K. Matsuura for the detailed information of the temperature and precipitation stations they have used.

## References

Ault, T.R., J.E. Cole, J.T. Overpeck, G.T. Pederson, S. St. George, B. Otto-Bliesner, B., C.A. Woodhouse and C. Deser (2013). The Continuum of Hydroclimate Variability in Western North America during the Last Millennium, J. Clim., 26, 16, 5863. DOI. https://doi.org/10.1175/JCLI-D-11-00732.1.

Blesić, S., D. Zanchettin and A. Rubino (2019). Heterogeneity of Scaling of the Observed Global Temperature Data, J. Clim., 32, 2, 349.

Bo, S., F. Schmidt, R. Eichhorn and G. Volpe (2019). Measurement of anomalous diffusion using recurrent neural networks, Phys. Rev. E, 100, 010102(R), https://doi.org/10.1103/PhysRevE.100.010102.

Bodri, L. (1994). Fractal analysis of climatic data: Mean annual temperature records in Hungary, Theor. Appl. Climatol. 49, 53-57.

Briane V, C. Kervrann and M. Vimond (2018). Statistical analysis of particle trajectories in living cells, Phys Rev E. Jun.; 97(6-1):062121., doi:10.1103/PhysRevE.97.062121. PMID: 30011544.

Bunde, A. and S. Havlin (2002). Power-law persistence in the atmosphere and in the oceans, Physica A, 314, 1-4, 15-24, https://doi.org/10.1016/S0378-4371(02)01050-6.

Bunde, A., J.F. Eichner, S. Havlin, E. Koscielny-Bunde, H.J. Schellnhuber and D. Vyushin (2004). Comment on "Scaling of Atmosphere and Ocean Temperature Correlations in Observations and Climate Models"., Phys. Rev. Lett. 92, 3, 039801, https://doi.org/10.1103/PhysRevLett.92.039801.

Bunde, A., U. Büntgen, J. Ludescher, J. Luterbacher and H. von Storch (2013). Is there memory in precipitation? Nat. Clim. Change, 3, 3, 174-175.

Caccia, D.C., D. Percival, M.J. Cannon, G. Raymond and J.B. Bassingthwaighte (1997). Analyzing exact fractal time series: evaluating dispersional analysis and rescaled range methods, Physica A, 246, 3-4, 609-632, https://doi.org/10.1016/S0378-4371(97)00363-4.

Chen, D. and H.W. Chen (2013). Using the Köppen classification to quantify climate variation and change: An example for 1901–2010, Environ. Develop., 6, 69–79.

Eichner, J.F., E. Koscielny-Bunde, A. Bunde, S. Havlin and H.J. Schellnhuber (2003). Power-law persistence and trends in the atmosphere: A detailed study of long temperature records, Phys.Rev. E, 68, 4, 046133, doi:10.1103/PhysRevE.68.046133.

Feder, J., (1988). Fractals, Plenum Press, New York.

Fraedrich, K. and R. Blender (2003). Scaling of Atmosphere and Ocean Temperature Correlations in Observations and Climate Models, Phys. Rev. Lett., 90, 10, 108501, https://doi.org/10.1103/PhysRevLett.90.108501.

Fraedrich, K. and C. Larnder (1993). Scaling regimes of composite rainfall time series, Tellus A, 45, 4, 289, DOI. https://onlinelibrary.wiley.com/doi/abs/10.1034/j.1600-0870.1993.t01-3-00004.x.

Garcin, M., (2017). Estimation of time-dependent Hurst exponents with variational smoothing and application to forecasting foreign exchange rates, Physica A, 483, 462, DOI:10.1016/j.physa.2017.04.122

Graves, T., R. Gramacy, N. Watkins, C. Franzke (2017). A Brief History of Long Memory: Hurst, Mandelbrot and the Road to ARFIMA, 1951-1980, Entropy, 19, 9, 437, https://doi.org/10.3390/e19090437

Grech, D. and Z. Mazur (2013). On the scaling ranges of detrended fluctuation analysis for long-term memory correlated short series of data, Physica A, 392, 10, 2384-2397, https://doi.org/10.1016/j.physa.2013.01.049.

Habib, A., J.P.R. Sorensen, J.P. Bloomfield, K. Muchan, A.J. Newell, A.P. Butler (2017). Temporal scaling phenomena in groundwater-floodplain systems using robust detrended fluctuation analysis, J. Hydrol. 549, 715-730, https://doi.org/10.1016/j.jhydrol.2017.04.034.

Hansen, J., R. Ruedy, M. Sato, and K. Lo (2010). Global surface temperature change, Rev. Geophys., 48, RG4004, doi:10.1029/2010RG000345.

Harris, D.C and P.D. Jones (2017). Time-Series (TS) version CRU 4.01 of high-resolution gridded data of month-by-month variation in climate (Jan. 1901- Dec. 2016).

Hurst, H.E. (1951). Long-term storage capacity of reservoirs. Trans. Am. Soc. Civ. Eng. 116, 1, https://doi.org/10.1061/TACEAT.0006518.

Janczura, J., P. Kowalek, H. Loch-Olszewska, J. Szwabiński, and A. Weron (2020). Classification of particle trajectories in living cells: Machine learning versus statistical testing hypothesis for fractional anomalous diffusion, Phys. Rev. E 102, 032402.

Jeon, J-H., E. Barkai and R. Metzler (2013). Noisy continuous time random walks, J. Chem. Phys., 139, 121916, https://doi.org/10.1063/1.4816635.

Kantelhardt, J.W., E. Koscielny-Bunde, H.H.A. Rego, S. Havlin and A. Bunde (2001). Detecting long-range correlations with detrended fluctuation analysis, Physica A, 295, 441-454, https://doi.org/10.1016/S0378-4371(01)00144-3.

Kantelhardt, J.W., S.A. Zschiegner, E. Koscielny-Bunde, S. Havlin., A. Bunde and H.E. Stanley (2002). Multifractal detrended fluctuation analysis of nonstationary time series, Physica A, 316, 1-4, 87-114,

https://doi.org/10.1016/S0378-4371(02)01383-3.

Kantelhardt, J.W., E. Koscielny-Bunde, D. Rybski, P. Braun, S. Havlin,and A. Bunde (2006). Long-term persistence and multifractality of precipitation and river runoff records, J. Geophys. Res., 111(D1), https://doi.org/10.1029/2005JD005881.

Kepten, E., A. Weron, G. Sikora, K. Burnecki and Y. Garini (2015). Guidelines for the Fitting of Anomalous Diffusion Mean Square Displacement Graphs from Single Particle Tracking Experiments, PLoS ONE, 10, 2, e0117722, https://doi.org/10.1371/journal.pone.0117722.

Király, A. and I.M. Jánosi (2005). Detrended fluctuation analysis of daily temperature records: Geographic dependence over Australia, Meteorol. Atmos. Phys., 88, doi:10.1007/s00703-004-0078-7.

Király, A., I. Bartos and I.M. Jánosi (2006). Correlation properties of daily temperature anomalies over land, Tellus A, 58, 593-600. https://doi.org/10.1111/j.1600-0870.2006.00195.x.

Koscielny-Bunde, E., A. Bunde, S. Havlin, H.E. Roman, Y. Goldreich and H.J. Schellnhuber (1998). Indication of a Universal Persistence Law Governing Atmospheric Variability. Phys. Rev. Lett., 81, 729, https://doi.org/10.1103/PhysRevLett.81.729.

Kottek, M., J. Grieser, C. Beck, B. Rudolf and F. Rubel, F (2006). World Map of the Köppen-Geiger climate classification updated, Meteorol. Zeit., 15, 259-263, doi:10.1127/0941-2948/2006/0130.

Kroese, D.P. and Z.I. Botev (2015). Spatial Process Simulation In: Schmidt V. (eds), Stochastic Geometry, Spatial Statistics and Random Fields, Lecture Notes in Mathematics, v. 2120, Springer, Cham.

Lanoiselée. Y., G. Sikora, A. Grzesiek, D.S. Grebenkov and A. Wyłomańska (2018). Optimal parameter for anomalous-diffusion-exponent estimation ferom noisy data, Phys. Rev. E 98, 062139.

Lenssen, N., G. Schmidt, J. Hansen, M. Menne, A. Persin,R. Ruedy and D, Zyss (2019). Improvements in the GISTEMP uncertainty model. J. Geophys. Res. Atmos., 124, 12, 6307-6326, doi:10.1029/2018JD029522.

Lenton, T.M., V. Dakos, S. Bathiany and M. Scheffer (2017). Observed trends in the magnitude and persistence of monthly temperature variability, Sci. Rep., 7, 5940, https://doi.org/10.1038/s41598-017-06382-x.

Mann, M.E., R.S. Bradley and N.K. Hughes (1998). Global-scale temperature patterns and climate forcing over the past six centuries., Nature, 392, 779-787, https://doi.org/10.1038/33859.

Markonis, Y. and D. Koutsoyiannis (2016). Scale-dependence of persistence in precipitation records, Nat. Clim. Change, 6,4, 399-401, https://doi.org/10.1038/nclimate2894

Monetti, R.A., S. Havlin and A. Bunde (2003). Long-term persistence in the sea surface temperature fluctuations, Physica A, 320, 581-589, doi:10.1016/S0378-4371(02)01662-X.

NOAA: 2020, https://www.ncdc.noaa.gov/cag/.

Peel, M.C., B.C. Finlayson and T.A. McMahon (2007). Updated world map of the Köppen-Geiger climate classification, Hydrol. Earth Syst. Sci., 11, 1633–1644.

Pelletier, J.D. and D.L. Turcotte (1997). Long-range persistence in climatological and hydrological time series: analysis, modeling and application to drought hazard assessment. J. Hydrol., 203, 198-208, https://doi.org/10.1016/S0022-1694(97)00102-9.

Rea, W., L. Oxley, M. Reale and J. Brown (2009). Estimators for Long Range Dependence: An Empirical Study. Electr. J. Statist., 0, https://arxiv.org/abs/0901.0762.

Robinson, P.J. (1995). Log-periodogram regression of time series with long-range dependence, Ann. Statist. 23, 1048-1072, doi:10.1214/aos/1176324636

Runge, C. (1901). Über empirische Funktionen und die Interpolation zwischen äquidistanten Ordinaten, Zeitschrift für Mathematik und Physik, 46, 224–243.

Shao, Z.G. and P.D. Ditlevsen (2016). Contrasting scaling properties of interglacial and glacial climates., Nat. Commun, 7, 10951, https://doi.org/10.1038/ncomms10951.

Sippel, S., J. Zscheischler, M. Heimann, F.E.L. Otto, J. Peters and M.D. Mahecha (2015). Quantifying changes in climate variability and extremes: Pitfalls and their overcoming, Geophys. Res. Lett., 42, 22, 9990-9998, https://doi.org/10.1002/2015GL066307.

Talkner, P. and R.O. Weber (2000). Power spectrum and detrended fluctuation analysis: Application to daily temperatures, Phys. Rev. E, 62, 150-160, doi:10.1103/PhysRevE.62.150.

Taqqu, M.S., V. Teverovsky and W. Willinger (1995). Estimators for long-range dependence: an empirical study, Fractals, 03, 04, 785.-798, https://doi.org/10.1142/S0218348X95000692.

Valverde, J.M., A. Castellanos and M.A.S. Quintanilla (2003). The memory of granular materials, Contemp. Phys., 44,

389, https://doi.org/10.1080/0010751031000155939

Varotsos, C., M.-N. Assimakopoulos and M. Efstathiou (2007). Technical Note: Long-term memory effect in the atmospheric CO2 concentration at Mauna Loa. Atmosph. Chem. Phys., 7, 3, 629-634, https://doi.org/10.5194/acp-7-629-2007

Varotsos, C.A., M.N. Efstathiou and A.P. Cracknell (2013). On the scaling effect in global surface air temperature anomalies, Atmosph. Chem.Phys, 13, 5243-5353, https://doi.org/10.5194/acp-13-5243-2013.

Weron, R. (2002). Estimating long-range dependence: finite sample properties and confidence intervals, Physica A, 312, 285-299, https://doi.org/10.1016/S0378-4371(02)00961-5.

Weron, A., J. Janczura, E. Boryczka, T. Sungkaworn and D. Calebiro (2019). Statistical testing approach for fractional anomalous diffusion classification, Phys. Rev. E, 99, 042149, https://doi.org/10.1103/PhysRevE.99.042149

Willmott, C.J. and K. Matsuura (2018). Terrestrial air temperature and precipitation: Monthly and annual time series (1900 - 2017), http://climate.geog.udel.edu/~climate/html_pages/download.html.

Yang, L. and Z. Fu (2019). Process-dependent persistence in precipitation records, Physica A, 527, 121459, https://doi.org/10.1016/j.physa.2019.121459

Zhang, W.F., Zhao, Q. (2015), Asymmetric long-term persistence analysis in sea surface temperature anomaly, Physica A, 428, 314-318, https://doi.org/10.1016/j.physa.2015.01.081.

***CORRESPONDING AUTHOR: Jouni J. TAKALO**,
University of Oulu and University of Jyväskylä,
Kontio, Finland
e-mail: jojuta@gmail.com